\documentclass[aps,prl,twocolumn,superscriptaddress]{revtex4-2}
\usepackage{gensymb}
\usepackage{color}
\usepackage[normalem]{ulem}
\usepackage{graphicx}
\usepackage{dcolumn}
\usepackage{float}
\usepackage{bm}
\usepackage{amsmath}
\usepackage{comment}

\newcommand{\be}{\begin{equation}}
\newcommand{\ee}{\end{equation}}
\newcommand{\bfig}{\begin{figure}}
\newcommand{\efig}{\end{figure}}

\newcommand{\LRSG}{LaRu$_3$(Si$_{1-x}$Ge$_x$)$_2$}
\newcommand{\LRS}{LaRu$_3$Si$_2$}

\newcommand{\RX}{\textit{R}$_8$Co\textit{X}$\kern-0.15em_{3}$}

\newcommand{\YRB}{YRu$_3$B$_2$}
\newcommand{\YRS}{YRu$_3$Si$_2$}

\begin{document}
\title{Bulk superconductivity in the kagome metal \YRB{}}
\author{Tobi Gaggl}
\affiliation{Department of Applied Physics and Quantum-Phase Electronics Center (QPEC), The University of Tokyo, Bunkyo, Tokyo 113-8656, Japan}
\affiliation{Physics Department, TUM School of Natural Sciences, Technical University of Munich, D-85748 Garching, Germany}

\author{Ryo Misawa}
\affiliation{Department of Applied Physics and Quantum-Phase Electronics Center (QPEC), The University of Tokyo, Bunkyo, Tokyo 113-8656, Japan}

\author{Markus Kriener}
\affiliation{RIKEN Center for Emergent Matter Science (CEMS), Wako, Saitama 351-0198, Japan}

\author{Yuki Tanaka}
\affiliation{Department of Applied Physics and Quantum-Phase Electronics Center (QPEC), The University of Tokyo, Bunkyo, Tokyo 113-8656, Japan}

\author{Rinsuke Yamada}
\affiliation{Department of Applied Physics and Quantum-Phase Electronics Center (QPEC), The University of Tokyo, Bunkyo, Tokyo 113-8656, Japan}

\author{Max Hirschberger}
\thanks{hirschberger@ap.t.u-tokyo.ac.jp}
\affiliation{Department of Applied Physics and Quantum-Phase Electronics Center (QPEC), The University of Tokyo, Bunkyo, Tokyo 113-8656, Japan}
\affiliation{RIKEN Center for Emergent Matter Science (CEMS), Wako, Saitama 351-0198, Japan}

\date{\today}

\begin{abstract}
Materials with a kagome lattice have been heavily studied recently for their exotic electronic band structure, structural frustration, high-temperature charge order transitions, and unconventional electron-phonon coupling. In \LRS{}, it was proposed that electronic flat bands conspire with the characteristic phonon spectrum of the kagome lattice to drive enhanced superconductivity at $T_c = 7\,$K. Here, we report bulk superconductivity in the structural analogue \YRB{}, which hosts a structurally pristine kagome lattice. We observe a superconducting transition at $T_c = 0.7\,$K through magnetization, resistivity, and heat-capacity measurements in this novel kagome metal.

\end{abstract}

\maketitle
%
%
%
%
%
%
%


\begin{figure}[htb]
  \begin{center}
		\includegraphics[clip, trim=0cm 0cm 0cm 0cm, width=0.98\linewidth]{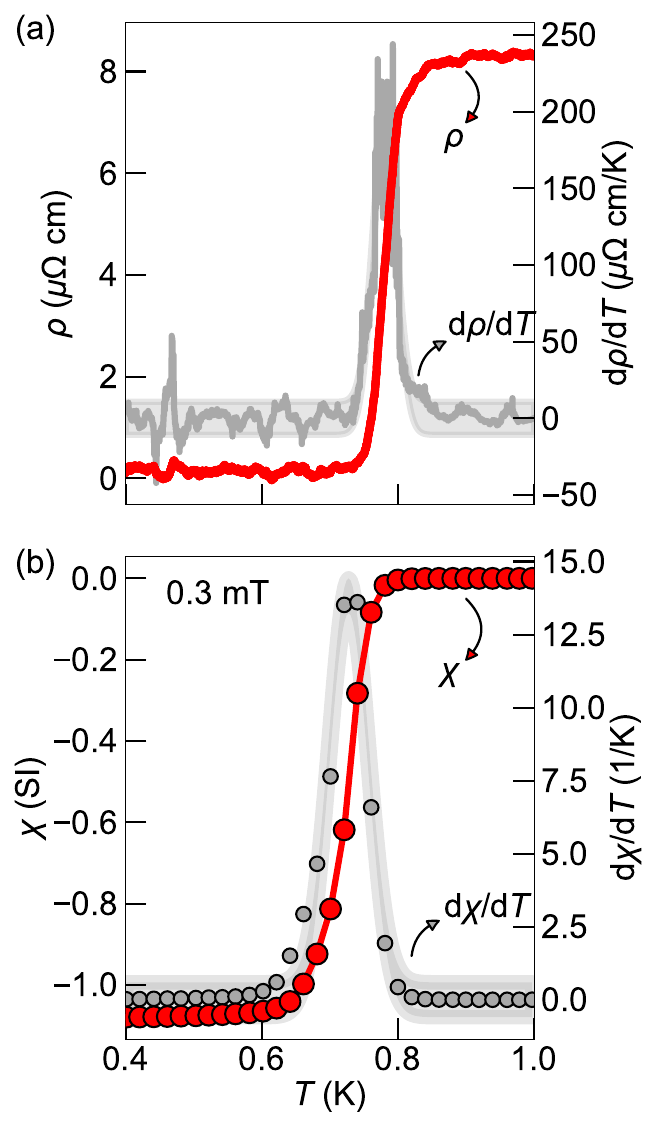}
    \caption[]{(color online). Superconducting transition in the kagome metal \YRB{}. (a) Resistivity $\rho_{xx}$ versus temperature $T$ measured on a bar-shaped sample in zero magnetic field, $H_\mathrm{ext} = 0$. Right axis: derivative $d\rho_{xx}/dT$ with respect to temperature. The transition width is $40\,$mK. (b) Magnetic susceptibility $\chi$ as a function of $T$ in $\mu_0H_\mathrm{ext} = 0.3\,$mT applied magnetic field, in zero-field cooled (ZFC) condition. Right axis: derivative $d\chi/dT$ with respect to temperature. The transition width is $70\,$mK. A demagnetization correction was applied to the $\chi(T)$ data.} 
    \label{Fig1}
  \end{center}
\end{figure}

\begin{figure}[htb]
  \begin{center}
    \includegraphics[clip, trim=0cm 0cm 0cm 0cm, width=1.\linewidth]{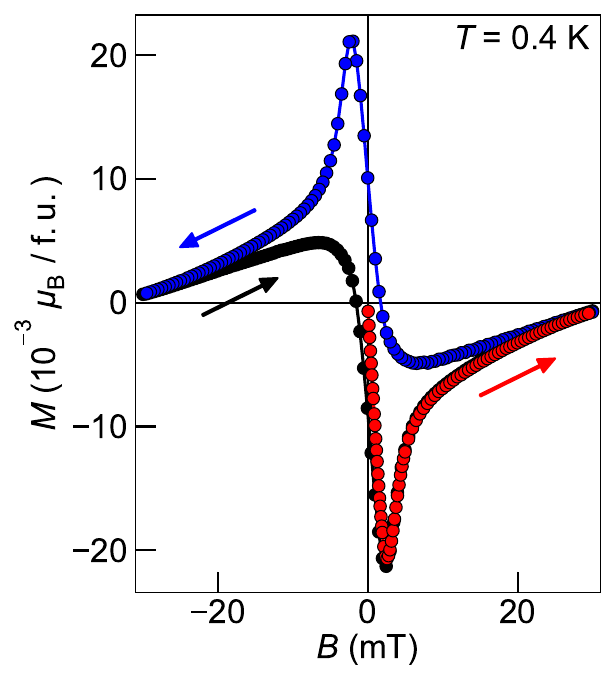}
    \caption[]{(color online). Perfect diamagnetism in \YRB{} detected by DC magnetometry. The magnetization isotherm $M(H)$ at base temperature, $T=0.4\,$K, shows a butterfly shape typical for type-II superconductors~\cite{Tinkham2004}. The critical field is $\mu_0H_{\mathrm{c2}}=30\,$mT. The curve is recorded in the sequence red (initial, zero-field cooled), blue ($\partial H /\partial t<0$), black ($\partial H / \partial t >0$).
    }
    \label{Fig2}
  \end{center}
\end{figure}


\begin{figure}[htb]
  \begin{center}
		\includegraphics[clip, trim=0cm 0cm 0cm 0cm, width=0.98\linewidth]{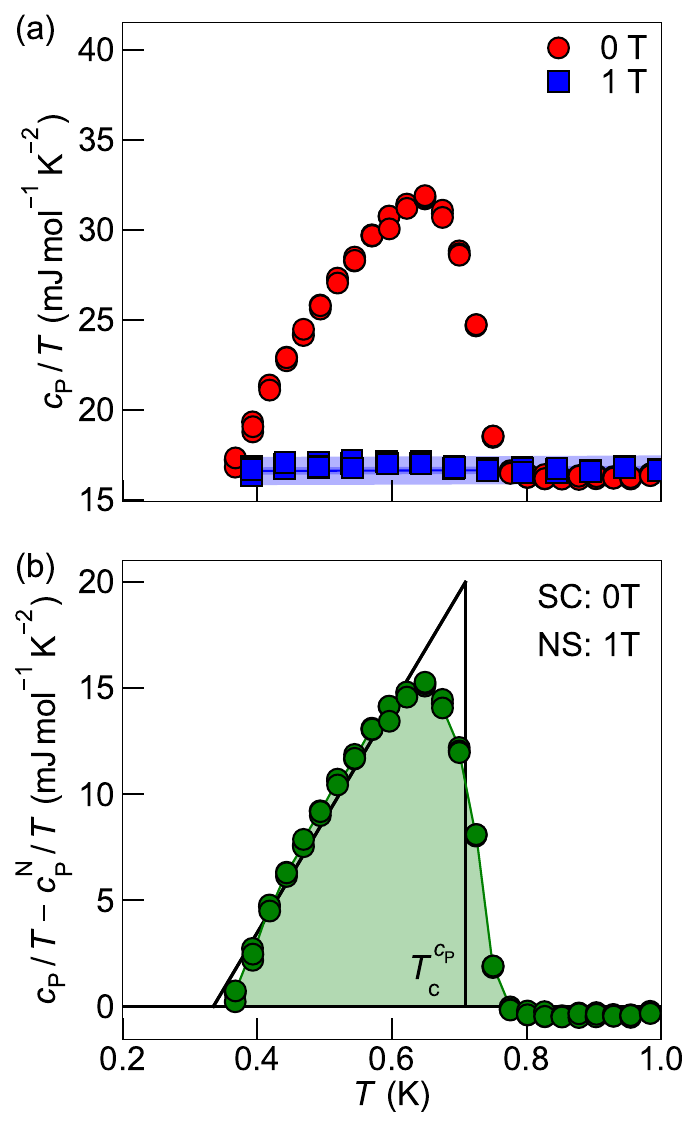}
    \caption[]{(color online). Thermodynamic evidence for superconductivity in \YRB{}. (a) Molar heat capacity at constant pressure, $c_\mathrm{P}$, at $B=0\,$T, $1\,$T for a polycrystalline sample of \YRB{}. The blue line indicates a polynomial fit to the high field data comprising electronic and phononic terms. (b) Superconducting part of $c_\mathrm{P}$ at $B=0\,$T. A fit to the BCS expression, with equal area construction for the anomaly in $c_\mathrm{P}$, is shown by the black line (see text for discussion). }
    \label{Fig3}
  \end{center}
\end{figure}


Among kagome metals, \LRS{} in the $RT_3M_2$ (1-3-2) family -- with $R$, $T$, $M$ being a rare earth (or actinide), transition metal, and main group element, respectively -- realizes the highest superconducting transition temperature, $T_c = 6.8\,$K~\cite{Barz1980}. \LRS{} has electronic flat bands with a high density of states close to the Fermi energy $E_\mathrm{F}$~\cite{Mielke2021}, and recent \textit{ab-initio} work by Deng~\textit{et al.} concluded that strong and mode-selective electron-phonon coupling drives the elevated superconducting $T_c$~\cite{Deng2025}. Ref.~\cite{Deng2025} also predicted that $T_c$ and electron-phonon coupling are highly tunable by chemical composition control, and some evidence for this scenario is obtained by the enhanced $T_c$ in the alloy series \LRSG{} and a concomitant analysis of the heat capacity~\cite{Misawa2025a}. 

There are further superconductors among the $R$Ru$_3$Si$_2$: \YRS{} and ThRu$_3$Si$_2$ have a moderately high $T_c$, $3.4\,$K~\cite{Barz1980,Gong2022} and $3.8\,$K~\cite{Barz1980,Liu2024}, respectively, and for ThRu$_3$Si$_2$ the reduction of $T_c$ as compared to \LRS{} was explained by a shift of the chemical potential, moving the kagome flat bands away from $E_\mathrm{F}$. CeRu$_3$Si$_2$ has a much lower $T_c\approx 1\,$K~\cite{Rauchschwalbe1984-wk, Rauchschwalbe1985}. For URu$_3$Si$_2$, no superconductivity has been reported~\cite{Rauchschwalbe1985}.
We also note that LuRu$_3$Si$_2$ could not be synthesized so far, including by us.

High-temperature structural instabilities and charge ordering has been well studied since the 1980s for \LRS{}~\cite{Vandenberg1980, Plokhikh2024, Mielke2025, Ma2025} and \YRS{}~\cite{Kral2025}. In our own recent work, we classified the 1-3-2 crystal structures using ionic electronegativity and a structural tolerance factor; we thus revealed the systematics of orthorhombic distortions of these kagome metals due to interplane coupling~\cite{Misawa2025b}. For example, the $R$Ru$_3$Si$_2$ are expected to all have a high-temperature orthorhombic instability of the hexagonal kagome network, whereas the $R$Ru$_3$B$_2$ do not. We thus thought it interesting to search for superconductors among the $R$Ru$_3$B$_2$ as a structural analogue to \LRS{} without high-temperature orthorhombic distortion. Previously, Ku~\textit{et al.} stabilized La$_{0.9}$Ru$_3$B$_2$, YRu$_3$B$_2$, and URu$_3$B$_2$ and found no superconductivity down to $1.2\,$K; in ThRu$_3$B$_2$, they reported $T_c=1.6\sim1.8\,$K~\cite{Ku1980}. Here, we take another look at \YRB{}, exploring the regime of lower temperature and observing a bulk superconducting transition at $T_c = 0.71\,$K.

\section{Experimental Methods}
\YRB{} crystals are grown from a stoichiometric combination of elements by arc-melting in an argon atmosphere after evacuation to $10^{-3}\,$Pa. First Ru and B were melted and turned 2 times, then Y was added and again turned 2 times. The ingots exhibit a shiny metallic luster. The weight loss during the melting was kept below $0.75\,\%$. The product is examined by powder X-ray diffraction, energy dispersive X-ray spectroscopy, and examination under a microscope equipped with a nicole prism. A Rietveld refinement is shown in Supplementary Fig.~\ref{Fig0}.

Resistivity measurements $\rho_{xx}(T)$ were carried out using a conventional four-probe method with the adiabatic demagnetization (ADR) option for a Quantum Design (QD) Physical Property Measurement System (PPMS) cryostat. The excitation current was $100\,\mathrm{\mu A}$. The data was recorded during the slow heating of the sample stage. 

We measured magnetization $M(T)$, $M(H)$ in a QD Magnetic Property Measurement System (MPMS-3) equipped with a $^3$He refrigerator (iQuantum Helium-3) for extraction magnetometry. 
The remanent field in the magnet was reduced as much as possible by carefully oscillating the field down to zero. Data were then taken after cooling in zero field to the base temperature for $M(T)$ in $\mu_0 H_\mathrm{ext} = 0.3\,$mT. Prior to each measurement, the temperature $T$ and external magnetic field $H_\mathrm{ext}$ were  stabilized.
The demagnetization correction is applied for the magnetization vs. temperature data: $\chi = \chi_\mathrm{raw} / (1+ N\chi_\mathrm{raw})$ with demagnetization factor $N\approx 1/3$ and $\chi_\mathrm{raw} = M / H_\mathrm{ext}$.

The molar heat capacity at constant pressure, $c_\mathrm{P}$, was measured by the relaxation technique in a QD PPMS equipped with the $^3$He refrigerator and $c_\mathrm{P}$ options. Again, the field was set to zero in oscillation mode. 
Addenda data were recorded prior the sample measurements at zero and finite fields and duly subtracted from the latter which was taken while
cooling the sample from $T > 2\,$K.

\section{Experimental Results}
The vanishing $\rho_{xx}$ of a bulk superconductor is its most technologically relevant physical property. In \YRB{}, the midpoint of the superconducting transition in the $\rho(T)$ is at $T_c^{\rho,\mathrm{mid}} =0.78\,$K with a full-width-half-maximum (FWHM) of $40\,$mK in $\partial \rho / \partial T$. The threshold of $5\,\%$ of the normal state resistance is reached at $T_c^{\rho,0} = 0.75\,$K. Above the superconducting transition, $\rho_{xx}$ follows a conventional metallic behavior with residual resistivity ratio $\rho_{xx}(300\,\mathrm{K})/ \rho(2\,\mathrm{K}) = 6$ for our arc-melted crystals. 

Figure~\ref{Fig1}(b) shows the magnetic susceptibility of \YRB{}. We confirm a superconducting shielding fraction of about $100\,\%$ ($\chi = –1$) after correcting for the demagnetization effect (see Experimental Methods). Data were taken upon warming in a field of $\mu_0 H = 0.3\,$mT after zero-field cooling to base temperature.
A sharp onset of diamagnetism, with a FWHM $70\,$mK, occurs at $T_c^m =0.73 \,$K. 
We next turn to the magnetic field dependence of the magnetization. In Fig.~\ref{Fig2}, $M(H)$ is shown at base temperature. The initial curve is linear at low field but deviates from linearity around $\mu_0 H_\mathrm{c1}\approx 2 \,$mT, subsequently decreasing towards zero at $\mu_0H_\mathrm{c2} = 30\,$mT. This data confirms that \YRB{} is a type-II superconductor. 

To ultimately confirm the bulk superconducting nature of \YRB{}, it is necessary to present a thermodynamic probe in zero magnetic field. We have measured $c_\mathrm{P}$ at $H = 0$ and $\mathrm{\mu}_0 H = 1\,$T, shown in Fig.~\ref{Fig3}(a). The zero-field data has a clear superconducting anomaly at $T_c^{c_\mathrm{P}}=0.71\,$K. For the analysis, the high-field data is first fitted to a polynomial expression $c_\mathrm{P}/T = \gamma + \beta T^2 + \eta T^4$ to account for the electronic ($\gamma$) and phononic ($\beta$, $\eta$) contributions to the heat capacity in the normal state of a metal. Next, we subtract $c_\mathrm{P}^\mathrm{sc} = c_\mathrm{P}(0\,\mathrm{T}) - c_\mathrm{P}(1\,\mathrm{T})$ and show the resulting data in Fig.~\ref{Fig3}(b). Using an equal area construction,
we fit $c_\mathrm{P}^\mathrm{sc}(T)$ and find $\Delta c_\mathrm{P} /(\gamma T_c)=1.30$, just below the BCS value of $1.43$.  We also observe a decrease of $c_\mathrm{P}$ at low temperatures, but our experimental range does not extend low enough to confirm the exponential dependence of $c_\mathrm{P}(T)$, which is expected for a fully gapped BCS superconductor when $T \ll T_c$. 

\section{Conclusion}
In this work, we identified superconductivity in the kagome metal \YRB{} with a transition temperature of $T_c = 0.7\,$K using magnetization, resistivity, and heat-capacity measurements. The relatively low $T_c$, especially when compared with the structural analogue \LRS{}, suggests that lattice effects and electronic structure may play different roles across this family of compounds. Our results provide insight into how lattice properties, charge-density-wave order, and superconductivity are connected in kagome metals, and they motivate further studies to clarify these interplays.

\textbf{Note added}: We became aware of a preprint on this compound~\cite{J-Winiarski2025-zm} after submission to a journal.

\section*{Acknowledgements}
This work was supported by JSPS KAKENHI Grants Nos. JP23H05431, JP24H01607, JP22K20348, JP23K13057, JP24H01604, JP25K17336 as well as JST CREST Grant Nos. JPMJCR1874 and JPMJCR20T1 (Japan), JST FOREST Grant No. JPMJFR2238 (Japan) and JST PRESTO Grant No. JPMJPR259A. This work was also supported by the Japan Science and Technology Agency (JST) as part of Adopting Sustainable Partnerships for Innovative Research Ecosystem (ASPIRE), Grant Number JPMJAP2426. M.H. is supported by the Deutsche Forschungsgemeinschaft (DFG, German Research Foundation) via Transregio TRR 360 – 492547816.

\bibliography{YRu3B2}

@ARTICLE{J-Winiarski2025-zm,
  title         = "{{YRu}$_$\_{3$}${B}$_$\_{2$}$ - a kagome lattice
                   superconductor}",
  author        = "J Winiarski, Michał and Walczak, Dominik and Królak, Szymon
                   and Yazici, Duygu and J Cava, Robert and Klimczuk, Tomasz",
  journal       = "arXiv:2512.08514 [cond-mat.supr-con]",
  month         =  "9~" # dec,
  year          =  2025,
  archivePrefix = "arXiv",
}

@article{Misawa2025a,
  title = {Chemical enhancement of superconductivity in ${\mathrm{LaRu}}_{3}{\mathrm{Si}}_{2}$ with mode-selective coupling between kagome phonons and flat bands},
  author = {Misawa, Ryo and Kriener, Markus and Yamada, Rinsuke and Nakano, Ryota and Jovanovic, Milena and Schoop, Leslie M. and Hirschberger, Max},
  journal = {Phys. Rev. Res.},
  volume = {7},
  issue = {3},
  pages = {033032},
  numpages = {11},
  year = {2025},
  month = {Jul},
  publisher = {American Physical Society},
  doi = {10.1103/rnv1-rbw2},
  url = {https://link.aps.org/doi/10.1103/rnv1-rbw2}
}

@article{Misawa2025b,
  title        = {{Successive orthorhombic distortions in kagome metals by molecular orbital formation}},
  author       = {Misawa, Ryo and Kitou, Shunsuke and Yamada, Rinsuke and Gaggl, Tobi and Nakano, Ryota and Shibata, Yudai and Okamura, Yoshihiro and Kriener, Markus and Nakamura, Yuiga and {\=O}nuki, Yoshichika and Takahashi, Youtarou and Arima, Taka-hisa and Jovanovic, Milena and Schoop, Leslie M. and Hirschberger, Max},
  journal      = {arXiv preprint arXiv:2507.17102},
  year         = {2025},
  archivePrefix= {arXiv},
  primaryClass = {cond-mat.str-el}
}

@article{Mielke2025,
  title        = {{Coexisting Multiple Charge Orders and Magnetism in the Kagome Superconductor LaRu3Si2}},
  author       = {Mielke, C. III and Sazgari, V. and Plokhikh, I. and Yi, Mingsheng and Shin, S. and Nakamura, H. and Graham, J. N. and Küspert, J. and Biało, I. and Garbarino, G. and Das, D. and Medarde, M. and Bartkowiak, M. and Yin, J.-X. and Islam, S. S. and Khasanov, R. and others},
  journal      = {Advanced Materials},
  year         = {2025},
  month        = {jul},
  doi          = {10.1002/adma.202503065},
  url          = {https://doi.org/10.1002/adma.202503065},
  note         = {First published: 23 July 2025}
}

@article{Plokhikh2024,
  title        = {{Discovery of charge order above room-temperature in the prototypical kagome superconductor La(Ru$_{1−x}$Fe$_x$)$_3$Si$_2$}},
  author       = {Plokhikh, I. and Mielke, C. III and Nakamura, H. and Petricek, V. and Qin, Y. and Sazgari, V. and Küspert, J. and Biało, I. and Shin, S. and Ivashko, O. and Graham, J. N. and Zimmermann, M. v. and Medarde, M. and Amato, A. and Khasanov, R. and Luetkens, H. and Fischer, M. H. and Hasan, M. Z. and Yin, J.-X. and Neupert, T. and Chang, J. and Xu, G. and Nakatsuji, S. and Pomjakushina, E. and Gawryluk, D. J. and Guguchia, Z.},
  journal      = {Communications Physics},
  year         = {2024},
  volume       = {7},
  pages        = {182},
  doi          = {10.1038/s42005-024-01625-4},
  url          = {https://doi.org/10.1038/s42005-024-01625-4}
}

@article{Ma2025,
  title        = {{Correlation between the dome-shaped superconducting phase diagram, charge order, and normal-state electronic properties in LaRu$_3$Si$_2$}},
  author       = {Ma, KeYuan and Plokhikh, Igor and Graham, Jennifer N. and Mielke, Charles III and Sazgari, Vahid and Nakamura, Hiroto and Islam, Shams Sohel and Shin, Soohyeon and Kr{\'a}l, Petr and Gerguri, Orion and Luetkens, Hubertus and von Rohr, Fabian O. and Yin, Jiaxin and Pomjakushina, Ekaterina and Felser, Claudia and Nakatsuji, Satoru and Wehinger, Bj{\"o}rn and Gawryluk, Dariusz J. and Medvedev, Sergey and Guguchia, Zurab},
  journal      = {Nature Communications},
  year         = {2025},
  volume       = {16},
  pages        = {6149},
  doi          = {10.1038/s41467-025-61184-y},
  url          = {https://doi.org/10.1038/s41467-025-61184-y}
}

@article{Kral2025,
  title        = {Discovery of High‐Temperature Charge Order and Time‐Reversal Symmetry‐Breaking in the Kagome Superconductor YRu$_3$Si$_2$},
  author       = {Kràl, P. and Graham, J. N. and Sazgari, V. and Plokhikh, I. and Lukovkina, A. and Gerguri, O. and Biało, I. and Doll, A. and Martinelli, L. and Oppliger, J. and Islam, S. S. and Wang, K. and Salamin, M. and Luetkens, H. and Khasanov, R. and Zimmermann, M. v. and Yin, J.-X. and Ziqiang Wang and Chang, J. and Monserrat, B. and Gawryluk, D. and von Rohr, F. O. and Kim, S.-W. and Guguchia, Z.},
  journal      = {arXiv preprint},
  year         = {2025},
  eprint       = {2507.06885},
  archivePrefix= {arXiv},
  primaryClass = {cond-mat.supr-con},
  url          = {https://arxiv.org/abs/2507.06885}
}

@article{Gong2022,
  title        = {{Superconductivity in Kagome Metal YRu$_3$Si$_2$ with Strong Electron Correlations}},
  author       = {Gong, Chunsheng and Tian, Shangjie and Tu, Zhijun and Yin, Qiangwei and Fu, Yang and Luo, Ruitao and Lei, Hechang},
  journal      = {Chinese Physics Letters},
  volume       = {39},
  number       = {8},
  pages        = {087401},
  year         = {2022},
  publisher    = {Chinese Physical Society and IOP Publishing Ltd},
  doi          = {10.1088/0256-307X/39/8/087401}
}

@article{Deng2025,
    title = {{Theory of Superconductivity in LaRu3Si2 and Predictions of New Kagome Flat Band Superconductors}},
    author = {Junze Deng and Yi Jiang and Tiago F. T. Cerqueira and Haoyu Hu and Eeli O. Lamponen and Dumitru Călugăru and Hanqi Pi, Zhijun Wang and Maia G. Vergniory and Emilia Morosan and Titus Neupert and S. Blanco-Canosa and Claudia Felser and Kristjan Haule and Miguel A. L. Marques and Päivi Törmä and B. Andrei Bernevig}, 
    journal = {arXiv:2503.20867},
    year = {2025},
}

@ARTICLE{Liu2024,
  title = "{{{{Superconductivity in kagome metal ThRu$_3$Si$_2$}}}}",
  author = "Yi Liu and Jing Li and Wu-Zhang Yang and Jia-Yi Lu and Bo-Ya Cao and Hua-Xun Li and Wan-Li Chai and Si-Qi Wu and Bai-Zhuo Li and Yun-Lei Sun and Wen-He Jiao and Cao Wang and Xiao-Feng Xu and Zhi Ren and Guang-Han Cao",
  journal       = "Chinese Physics B",
  volume        = 33,
  number        = 5,
  page          = 057401,
  year          =  2024,
}

@book{Tinkham2004,
  title = "{Introduction to superconductivity}",
  author = "M. Tinkham",
  year = "2004",
  publisher = "Courier Corporation",
}

@ARTICLE{Ku1980,
  title = "{{{{Superconducting and magnetic properties of new ternary borides with the CeCo$_3$B$_2$-type structure}}}}",
  author = "H. C. Ku and G. P. Meisner and F. Acker and D. C. Johnston",
  journal   = "Solid State Commun.",
  publisher = "Elsevier BV",
  volume    =  35,
  number    =  2,
  pages     = "91--96",
  month     =  "1~" # jul,
  year      =  1980
}

@ARTICLE{Rauchschwalbe1984-wk,
  title = "{{{{Superconductivity in a mixed-valent system: CeRu$_3$Si$_2$}}}}",
  author = "U. Rauchschwalbe and W. Lieke and F. Steglich and C. Godart and L. C. Gupta and R. D. Parks",
  journal   = "Phys. Rev. B Condens. Matter",
  publisher = "American Physical Society (APS)",
  volume    =  30,
  number    =  1,
  pages     = "444--446",
  month     =  "1~" # jul,
  year      =  1984
}

@ARTICLE{Rauchschwalbe1985,
  title = "{{{{Physical properties, electronic properties Investigation of new lanthanum-, cerium- and uranium-based ternary intermetallics☆}}}}",
  author = "U. Rauchschwalbe and U. Gottwick and U. Ahlheim and H.M Mayer and F. Steglich",
  journal   = "Journal of the Less Common Metals",
  volume    =  111,
  issue    =  "1–2",
  pages     = "265--275",
  year      =  1985
}

@ARTICLE{Barz1980,
  title = "{{{{New ternary superconductors with silicon}}}}",
  author = "H. Barz",
  journal = "Mater. Res. Bull.",
  volume  =  15,
  number  =  10,
  pages   = "1489--1491",
  month   =  "1~" # oct,
  year    =  1980
}

@ARTICLE{Mielke2021,
  title = "{{{{Nodeless kagome superconductivity in {LaRu}$_{3}${Si}$_{2}$}}}}",
  author = "C. Mielke and Y. Qin and J. Yin and H. Nakamura and D. Das and K. Guo and R. Khasanov and J. Chang and Z. Q. Wang and S. Jia and S. Nakatsuji and A. Amato and H. Luetkens and G. Xu and M. Z. Hasan and Z. Guguchia",
  journal   = "Phys. Rev. Mater.",
  publisher = "American Physical Society",
  volume    =  5,
  number    =  3,
  pages     =  034803,
  month     =  "29~" # mar,
  year      =  2021
}

@article{Vandenberg1980,
  title        = {The crystal structure of a new ternary silicide in the system rare-earth--ruthenium--silicon},
  author       = {Vandenberg, J. M. and Barz, H.},
  journal      = {Materials Research Bulletin},
  year         = {1980},
  volume       = {15},
  pages        = {1493--1498},
  doi          = {10.1016/0025-5408(80)90565-2},
  issn         = {0025-5408},
  publisher    = {Pergamon Press},
  address      = {New York, USA}
}

\clearpage
\setcounter{page}{1}
\begin{widetext}
\begin{center}
\Large{Supplementary Information}
\end{center}

Fig.~\ref{Fig0} shows the powder X-ray diffraction pattern of crushed \YRB{} crystals, together with a Rietveld fit in the $P6/mmm$ space group without adjustable parameters for the internal coordinates of the constituent atoms. The lattice constants at room temperature are $a = 5.4706\,$\AA, $c = 3.0275\,$\AA. There are no impurity peaks at the resolution of $1\,\%$ of the dominant main phase peak. 

\begin{figure}[h]
  \begin{center}
		\includegraphics[clip, trim=0cm 0cm 0cm 0cm, width=0.5\linewidth]{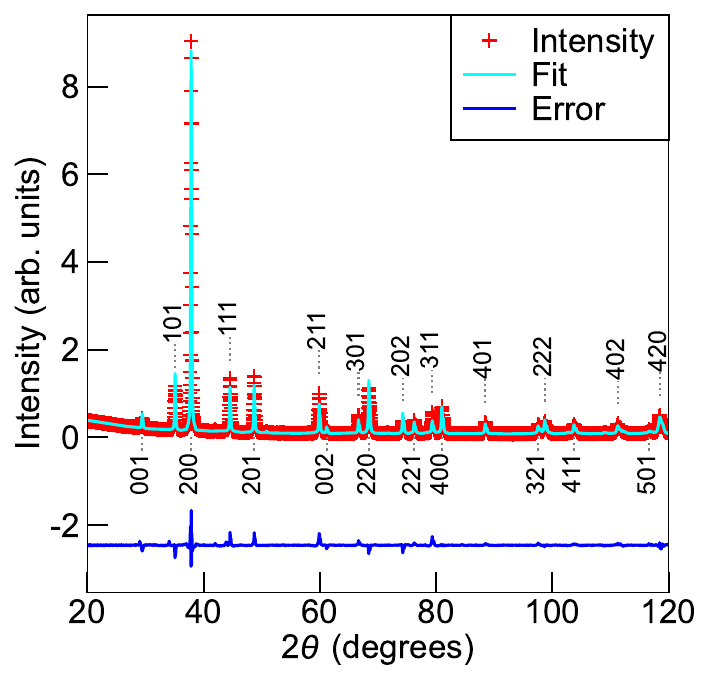}
    \caption[]{(color online). Rietveld refinement of the crystal structure of \YRB{}, based on a crushed polycrystalline sample. The powder X-ray data was obtained on a commercial Rigaku SmartLab diffractometer with Cu-$K_\alpha$ radiation (wavelength $\lambda = 1.5406 \, \AA$).} 
    \label{Fig0}
  \end{center}
\end{figure}

\end{widetext}
\end{document}